\documentstyle[12pt,epsfig]{article}
\newlength{\dinwidth}
\newlength{\dinmargin}
\newlength{\extraspace}
\newlength{\extraspaces}
\newcommand{\be}{\begin{equation}
\addtolength{\abovedisplayskip}{\extraspaces}
\addtolength{\belowdisplayskip}{\extraspaces}
\addtolength{\abovedisplayshortskip}{\extraspace}
\addtolength{\belowdisplayshortskip}{\extraspace}}
\newcommand{\ee}{\end{equation}}
\newcommand{\bdm}{\begin{displaymath}
\addtolength{\abovedisplayskip}{\extraspaces}
\addtolength{\belowdisplayskip}{\extraspaces}
\addtolength{\abovedisplayshortskip}{\extraspace}
\addtolength{\belowdisplayshortskip}{\extraspace}}
\newcommand{\edm}{\end{displaymath}}
\renewcommand{\thefootnote}{\fnsymbol{footnote}}
\def\simlt{\mathrel{\lower2.5pt\vbox{\lineskip=0pt\baselineskip=0pt
           \hbox{$<$}\hbox{$\sim$}}}}
\def\simgt{\mathrel{\lower2.5pt\vbox{\lineskip=0pt\baselineskip=0pt
           \hbox{$>$}\hbox{$\sim$}}}}
\newcommand{\prd}{{\em Phys.\ Rev.\ }  {\bf D}}

\newcommand{\prl}{{\em Phys.\ Rev.\ Lett.\ }}
\newcommand{\np}{{\em Nucl.\ Phys.\ }{\bf B}}
\newcommand{\pl}{{\em Phys.\ Lett.\ }{\bf B}}

\newcommand{\zp}{Z. Phys.\ {\bf C}}

\begin{document}
%
\setcounter{footnote}{1}
\vspace{6mm}
\begin{center}
\Large{{\bf Anomalous Couplings of the Third Generation\\
in Rare \boldmath{$B$} Decays}}
\end{center}
\vspace{5mm}
\begin{center}
{\large Gustavo Burdman $^{(1)}$, M.\ C.\ Gonzalez--Garcia
$^{(2,3)}$, and S.\ F.\ Novaes $^{(3)}$}

\vspace{10mm}
{\normalsize\it $^{(1)}$ Department of Physics, University of Wisconsin,}\\ 
{\normalsize\it Madison, WI 53706, USA.}\\

\vspace{7mm}
{\normalsize\it $^{(2)}$ Instituto de F\'{\i}sica Corpuscular -- IFIC/CSIC, 
Departament de  F\'{\i}sica Te\`orica \\
Universitat de Val\`encia, 46100 Burjassot, Val\`encia, Spain.}\\

\vspace{7mm}
{\normalsize\it $^{(3)}$ Instituto de F\'{\i}sica Te\'orica, 
Universidade Estadual Paulista \\
Rua Pamplona, 145, 01405--900  S\~a{o} Paulo, Brazil.} 

\end{center}

\vspace{0.50cm}
\thispagestyle{empty}
\begin{abstract}
We study the potential effect of anomalous couplings of the third
generation quarks to gauge bosons in rare $B$ decays.  We focus
on the constraints from flavor changing neutral current processes
such as $b\to s\gamma$ and $b\to s \ell^+\ell^-$.  We consider
both dimension-four and dimension-five operators and show that
the latter can give large deviations from the standard model in
the still unobserved dilepton modes, even after the bounds from
$b\to s\gamma$ and precision electroweak observables  are taken
into account. 
\end{abstract}
\newpage

\renewcommand{\thefootnote}{\arabic{footnote}}
\setcounter{footnote}{0}
\setcounter{page}{1}
\section{Introduction}  

The continuing experimental success of the standard model (SM)
suggests the possibility that additional particles and/or
non-standard interactions may only be found at scales much larger
than $M_W$. On the other hand, several questions remain
unanswered within the SM framework that may require new dynamics
in order to be addressed. Chief among these questions are the
origin of electroweak symmetry breaking and of fermion masses. In
principle, it could be argued that the energy scales of the new
dynamics related to these questions may be so large as to be
irrelevant to observables at the electroweak scale. However, it
is known that the physics behind the Higgs sector, responsible
for the breaking of the electroweak symmetry, cannot reside at
scales much higher than few TeV. Furthermore, it is possible that
the origin of the top quark mass might be related to electroweak
symmetry breaking.  Thus, at least in some cases, the dynamics
associated with new physics may not reside at arbitrarily high
energies and there might be some observable effects at lower
energies.

The effects of integrating out the physics residing at some high
energy scale $\Lambda\gg M_W$, can be organized in an effective
field theory for the remaining degrees of freedom. Such a theory
for the electroweak symmetry breaking (EWSB) sector of the SM
involves the electroweak gauge bosons as well as the
Nambu-Goldstone bosons (NGB) associated with the spontaneous
breaking of $SU(2)_L\times U(1)_Y$ down to $U(1)_{\rm
EM}$~\cite{longhitano,feruglio}. The effective theory must be
studied up to next-to-leading order for the possible departures
from the SM to appear. This program resembles that of chiral
perturbation theory for pions in low energy QCD where, for
instance, the presence of the $\rho$ resonance results in
deviations from the low energy theorems. For the case of the
electroweak interactions, a variety of electroweak precision
measurements and flavor changing neutral current processes
provide testing ground for possible deviations originating in the
EWSB sector of the SM. The next-to-leading order terms in the
effective theory will generally contribute to oblique
corrections, triple and quartic anomalous gauge boson couplings,
and corrections to the NGB propagators that result in
four-fermion interactions~\cite{pich}.

In addition to the low energy description of the interactions of
the EWSB sector (i.e. gauge bosons plus NGBs) one may consider
the possibility that the new physics above $\Lambda$ may also
modify the effective interactions of the SM fermions to the
electroweak gauge bosons. In principle, this also has a parallel
in low energy QCD, as it is pointed out in Ref.~\cite{peccei},
where symmetry alone is not enough to determine the axial
coupling of nucleons to pions. In fact, the departure of this
coupling from unity is a non-universal effect, only determined by
the full theory of QCD.  Thus, in Ref.~\cite{peccei} it is
suggested that in addition to the effects in the EWSB sector of
the theory, it is possible that the interactions of fermions with
the NGBs are affected by the new dynamics above $\Lambda$,
resulting in anomalous interactions with the electroweak gauge
bosons. This is particularly interesting if fermion masses are
dynamically generated, as is the case with the nucleon mass.
Interestingly, the proximity of the top quark mass to the
electroweak scale $v=246~$GeV, hints the possibility the top mass
might be a dynamically generated ``constituent'' mass.  Thus, it
is of particular interest to study the couplings of third
generation quarks to electroweak gauge bosons.

Processes involving FCNC transitions in $B$ and $K$ decays are a
crucial complement to precision electroweak observables, when
constraining the physics of the EWSB sector.  The effects of
anomalous triple gauge boson couplings~\cite{atgc}, as well as of
the corrections to NGB propagators~\cite{bews} give in each case
a distinct pattern of deviations from the SM expectations in rare
$B$ and $K$ decays.  On the other hand, the anomalous couplings
of third generation quarks to the $W$ and the $Z$ can come from
dimension-four and dimension-five operators.  The indirect effects of
the dimension-four operators have been considered in relation to
electroweak observables in Ref.~\cite{peccei,yuan}, as well as
the $b\to s\gamma$ transitions~\cite{fuji}.  The constraints on
dimension-five operators from electroweak physics have been studied
in Ref.~\cite{ewd5}.  In this paper, we consider the effects of
all possible dimension-five operators in $B$ FCNC transitions such
as $b\to s\gamma$ and $b\to s\ell^+\ell^-$.  For completeness, we
also present the analysis of the dimension-four operators.  We discuss
that with the very natural assumption of chiral symmetry, in fact
enforcing vanishing fermion mass renormalization in the chiral
limit, the effects of dimension-four operators found in
Ref.~\cite{fuji} for $b\to s\gamma$ are not so dramatic.
Moreover, we will see that the effects of dimension-five operators
are comparable and may even dominate over the supposedly leading
lower dimension contributions.

In Section~2 we present a brief introduction to the effective
theory approach and set our notation.  We present the constraints
from rare $B$ decays on the coefficients of dimension-four operators
in Section~3. In Section~4 we discuss the possible effects in
rare $B$ decays from all possible dimension-five operators involving
the third generation quarks.  Finally, we discuss the results and
conclude in Section~5.

\section{The Effective Theory}
 
If the Higgs boson, responsible for the electroweak symmetry
breaking, is very heavy, it can be effectively removed from the
physical low--energy spectrum. In this case and for dynamical
symmetry breaking scenarios relying on new strong interactions,
one is led to consider the most general effective Lagrangian
which employs a nonlinear representation of the spontaneously
broken $SU(2)_L \times U(1)_Y$ gauge symmetry \cite{appelquist}.
The resulting chiral Lagrangian is a non--renormalizable
nonlinear $\sigma$--model coupled in a gauge--invariant way to
the Yang-Mills theory.  This model-independent approach
incorporates by construction the low--energy theorems \cite{cgg}
that predict the general behavior of Goldstone boson amplitudes,
irrespective of the details of the symmetry breaking mechanism.
Unitarity requires that this low--energy effective theory should
be valid up to some energy scale smaller than $4\pi v \simeq 3$
TeV, where new physics would come into play.

In order to specify the effective Lagrangian for the Goldstone
bosons, we assume that the symmetry breaking pattern is $G =
SU(2)_L \times U(1)_Y$ $\longrightarrow$ $H = U(1)_{\mbox{em}}$,
leading to just three Goldstone bosons $\pi^a$ ($a=1,2,3$). With
this choice, the building block of the chiral Lagrangian is the
dimensionless unimodular matrix field $\Sigma$,
\begin{equation}
\Sigma = \exp \left(i\frac{\pi^a\tau^a}{v}\right) \; ,
\end{equation}
where $\tau^a$ ($a=1,2,3$) are the Pauli matrices. We implement
the $SU(2)_C$ custodial symmetry by imposing a unique
dimensionful parameter, $v$, for charged and neutral fields.
Under the action of $G$ the transformation of $\Sigma$ is
\[
\Sigma \rightarrow \Sigma' = L~ \Sigma~ R^\dagger \; ,
\]
where $L= \exp (i \alpha^a\tau^a/2 )$ and  $R = \exp
(iy\tau^3/2)$, with $\alpha^{a}$ and $y$ being the parameters of
the transformation.

The gauge fields are represented by the matrices $\hat{W}_{\mu} =
\tau^a W^a_{\mu}/(2i)$, $\hat{B}_{\mu} = \tau^3 B_{\mu}/(2i)$, while
the associated field strengths are given by
\begin{eqnarray*}
\hat{W}_{\mu\nu} &=& \partial_{\mu}\hat{W}_{\nu} - 
\partial_{\nu}\hat{W}_{\mu} -g\left[\hat{W}_{\mu},\hat{W}_{\nu}\right]\; ,
\\
\hat{B}_{\mu\nu} &=& \partial_{\mu}\hat{B}_{\nu} - \partial_{\nu}
\hat{B}_{\mu} \; .
\end{eqnarray*}
In the nonlinear representation of the gauge group $SU(2)_L
\times U(1)_Y$, the mass term for the vector bosons is given by
the lowest order operator involving the matrix $\Sigma$.
Therefore, the kinetic Lagrangian for the gauge bosons reads
\begin{equation}
{\cal L}_B = \frac{1}{2} {\rm Tr} \left( \hat{W}_{\mu\nu}
\hat{W}^{\mu\nu} + \hat{B}_{\mu\nu} \hat{B}^{\mu\nu}\right) +
\frac{v^2}{4}{\rm Tr}\left(D_{\mu}
\Sigma^{\dagger}D^{\mu}\Sigma\right)\;,
\label{mass}
\end{equation}
where the covariant derivative of the field $\Sigma$ is 
$ D_{\mu}\Sigma = \partial_{\mu}\Sigma - g\hat{W}_{\mu}\Sigma 
+ g' \Sigma \hat{B}_{\mu}$.

The effects of new dynamics on the couplings of fermions with
the SM gauge bosons can be, in principle,  also studied in an
effective Lagrangian approach.  For instance, if in analogy
with the situation in QCD, fermion masses are dynamically
generated in association with EWSB, residual interactions of
fermions with Goldstone bosons could be important~\cite{peccei}
if the $m_f\simeq f_{\pi}\simeq v$. Thus residual, non-universal
interactions of the third generation quarks with gauge bosons
could carry interesting information about both the origin of the
top quark mass and EWSB. 

In order to include fermions in this framework, we must define
their transformation under $G$.  Following Ref.\ \cite{peccei},
we postulate that matter fields feel directly only the
electromagnetic interaction $f \rightarrow f' = \exp(iyQ_f)\;f $,
where $Q_f$ stands for the electric charge of fermion $f$. The
usual left--handed fermion doublets are then defined with the
following transformation under $G$,  
\begin{equation}
\Psi_L = \Sigma \left(\begin{array}{c}f_1\\[-0.1cm] 
f_2\end{array}\right)_L 
\;\;\;\; \longrightarrow \;\;\;\; \Psi^\prime_L = L~ \exp(i y Y /2) \Psi_L ,
\end{equation}
where $Q_{f_1} - Q_{f_2} = 1$ and $Y = 2 Q_{f_1} -1$.
Right--handed fermions are just the singlets $f_R$. 
This definition is useful since it permits the construction of
linearly realized left-handed doublet fields in the same way that, 
when studying the breaking of 
$SU(2)_R \times SU(2)_L \rightarrow SU(2)_{R+L}$ in QCD,
one introduces auxiliary fields for the nucleons which transform
linearly under the broken axial group. 
In this framework, the lowest--order interactions between fermions and
vector bosons that can be built are of dimension four, leading to
anomalous vector and axial--vector couplings, which were analyzed
in detail in Ref.\ \cite{yuan}.

In order to construct the most general Lagrangian describing
these interactions, it is convenient to define the vector and
tensor fields
\begin{eqnarray}
\Sigma_{\mu}^a & = & -\frac{i}{2}
{\rm Tr}\left(\tau^a V_\mu^R\right) 
~=~ -\frac{i}{2}{\rm Tr}
\left(\tau^a\Sigma^{\dagger} D_{\mu}\Sigma\right) \; ,
\nonumber \\
\Sigma_{\mu\nu}^a & = & -i\;{\rm Tr}\left[\tau^a \Sigma^{\dagger}
\left[D_{\mu},D_{\nu}\right]\Sigma\right] \; .
\end{eqnarray}
Under $G$ transformations $\Sigma_{\mu}^3$ and $\Sigma_{\mu\nu}^3$ are
invariant while 
\[
\Sigma_{\mu(\mu\nu)}^{\pm} \rightarrow
{\Sigma'}_{\mu(\mu\nu)}^{\pm} = \exp (\pm iy) 
\Sigma_{\mu(\mu\nu)}^{\pm} \; ,
\]
where $\Sigma_{\mu(\mu\nu)}^{\pm} = (\Sigma_{\mu(\mu\nu)}^1 \mp
i\Sigma_{\mu(\mu\nu)}^2)/\sqrt{2}$.

The basic fermionic elements for the construction of neutral-
and charged-current effective interactions are
\begin{eqnarray}
 \Delta_X (q, q^\prime) & = & \bar{q} \; P_X \; q^\prime \; ,
\nonumber \\
\Delta^\mu_X(q,q') & = &\bar q \gamma^\mu \; P_X \; q' \; , 
\nonumber \\
\tilde\Delta^{\mu}_X (q, q^\prime) & = & \bar{q} \; P_X \; \tilde D^\mu
 q^\prime\; , 
 \\
\Delta^{\mu\nu}_X (q, q^\prime) &= &\bar{q} \sigma^{\mu\nu} \; P_X \;
q^\prime   \; , \nonumber
\label{fer}
\end{eqnarray}
where $P_X$ ($X=0$, $5$, $L$, and $R$) stands for $I$,
$\gamma^5$, $P_L$, and $P_R$ respectively, with $I$ being the
identity matrix and $P_{L(R)}$ the left (right) chiral projector.
The fermionic field $q$ ($q^\prime$) represents any quark flavor,
and $\tilde D^\mu$ stands for the electromagnetic covariant
derivative.

The most general dimension-four Lagrangian invariant under
nonlinear transformations under $G$ is
\begin{eqnarray}
{\cal L}_{{4}} & = & 
d_L^{{NC}} ~\Delta^\mu_L(t,t)~\Sigma_\mu^3 +
d_R^{{NC}} ~\Delta^\mu_R(t,t)~\Sigma_\mu^3 +
d_L^{CC}~\Delta^\mu_L(t,b)\Sigma^+_\mu 
\nonumber \\
& + &
d_L^{CC\dagger}\Delta^\mu_L(b,t)\Sigma^-_\mu + 
d_R^{CC}\Delta^\mu_R(t,b)\Sigma^+_\mu +
d_R^{CC\dagger}\Delta^\mu_R(b,t)\Sigma^-_\mu \; .
\label{d4}
\end{eqnarray}
In principle it is also possible to construct neutral current
operators involving only the bottom quark. We will assume
however, that these vertices are not modified by the dynamics of
the symmetry breaking or, at most, that these modifications are
suppressed as compared to those of the top quark. 

In a very general parameterization, the dimension-four anomalous
couplings of third generation quarks can be written in terms of
the usual physical fields as,
\begin{eqnarray}
{\cal L}_{{4}} &=&-\frac{g}{\sqrt{2}}\left[ C_L\;(\bar t_L\gamma_\mu b_L)
+C_R\;(\bar t_R\gamma_\mu b_R) \right] W^{+\mu} \nonumber\\
&& -\frac{g}{2\;c_W}\left[ N_L^t \;(\bar t_L\gamma_\mu t_L) 
+ N_R^t \;(\bar t_R\gamma_\mu t_R)\right]Z^\mu  + \mbox{h.c.} ~ , 
\label{lfer}
\end{eqnarray} 
where $s_W$ ($c_W$) is the sine (cosine) of the
weak mixing angle, $\theta_W$. The parameters $C_{L,R}$,
$N_{L,R}^{t}$ can be written in terms of the constants
$d_{L,R}^{NC,CC}$ of Eq.\ (\ref{d4}) and contain the residual,
non-universal effects associated with the new dynamics, perhaps
responsible for the large top quark mass. Then, if we assume that
the new couplings are CP conserving~\cite{german}, there are four
new parameters. They are constrained at low energies by a variety
of experimental information, mostly from electroweak precision
measurements and the rate of $b\to s\gamma$. 

In the case of dimension-five operators, the most general
neutral--current interactions, which are invariant under
nonlinear transformations under $G$, are \cite{yuan2},
\begin{eqnarray}
{\cal L}_5^{NC}  & = & a_1^{{NC}} ~\Delta_0(t,t)~  
\Sigma_\mu^+  \Sigma^{- \mu} 
+ a_2^{{NC}} ~ \Delta_0(t,t)~  \Sigma_\mu^3 \Sigma^{3 \mu}
+ i~ a_3^{{NC}} ~ \Delta_5(t,t)~\partial^\mu \Sigma^{3}_\mu 
\nonumber \\ 
& + & i~ b_1^{{NC}} ~ \Delta^{\mu\nu}_0 (t,t) ~
\mbox{Tr} \left [ T \hat{W}_{\mu\nu} \right ] 
+ b_2^{{NC}} ~ \Delta^{\mu\nu}_0 (t,t)~ B_{\mu\nu}
 \\ 
& + & i~ b_3^{{NC}}~ \Delta^{\mu\nu}_0 (t,t)~ 
\left ( \Sigma^+_\mu \Sigma^-_\nu - \Sigma^+_\nu \Sigma^-_\mu
\right )~
 + i~ c_{1}^{{NC}}~\left[\tilde\Delta^{\mu}_0 (t,t)
-\overline{\tilde\Delta^{\mu}_0}
(t,t)\right]\Sigma^{3 \mu} \; , \nonumber
\label{nc}
\end{eqnarray}
and the charged--current interactions are
\begin{eqnarray}
{\cal L}_5^{CC} & = & a_{1L}^{{CC}}~ \Delta_L(t,b)~  
\Sigma_\mu^+ \Sigma^{3 \mu} 
+ a_{1R}^{{CC}} ~ \Delta_R(t,b)~  \Sigma_\mu^+ \Sigma^{3 \mu} 
\nonumber \\
& + &i a_{2L}^{{CC}}~ \Delta_L(t,b)~  
\tilde D^\mu  \Sigma_\mu^+ 
+i a_{2R}^{{CC}} ~ \Delta_R(t,b)~  \tilde D^\mu \Sigma_\mu^+ 
\nonumber  \\  
& + & b_{1L}^{{CC}} \Delta^{\mu\nu}_L (t,b) ~
\Sigma_{\mu\nu}^+ 
 + \, b_{1R}^{{CC}}~ \Delta^{\mu\nu}_R (t,b) ~
\Sigma_{\mu\nu}^+ 
\label{cc} \\
& + &  b_{2L}^{{CC}}~ \Delta^{\mu\nu}_L (t,b)~ 
\left ( \Sigma^+_\mu \Sigma^3_\nu - \Sigma^+_\nu \Sigma^3_\mu
\right ) 
+\, b_{2R}^{{CC}}~ \Delta^{\mu\nu}_R (t,b)~ 
\left ( \Sigma^+_\mu \Sigma^3_\nu - \Sigma^+_\nu \Sigma^3_\mu
\right ) 
\nonumber \\
& + &i~ c_{1L}^{{CC}}~\tilde\Delta^{\mu}_L (t,b)\Sigma^{+}_\mu 
+ i~ c_{1R}^{{CC}}~\tilde\Delta^{\mu}_R (t,b)\Sigma^{+}_\mu 
~+~ \mbox{h.c.} \; . \nonumber 
\end{eqnarray}
In general, since chiral Lagrangians are related to strongly
interacting theories, it is hard to make firm statements about
the expected order of magnitude of the couplings.
Notwithstanding, requiring the loop corrections to the effective
operators to be of the same order of the operators themselves
suggests that these coefficients are of ${\cal O}(1)$
\cite{wudka}. Moreover, if the high energy theory respects chiral
symmetry, we can also foresee a further suppression factor
proportional to $m_t/\Lambda$.  

In the unitary gauge, we can rewrite the interactions (\ref{nc})
and (\ref{cc}) as a scalar, a vector, and a tensorial Lagrangian
involving the physical fields. For the Lagrangian involving
scalar currents we have,
\begin{eqnarray}
{\cal L}_S &=& \frac{g^2}{2\Lambda}
\Biggl[   \; \bar{t}\; t \; \Bigl(
\alpha_1^{NC} W^+_\mu W^{-\mu} + \frac{\alpha_2^{NC}}{2 c_W^2} Z^\mu Z_\mu
\Bigr) \Biggr] + i \frac{g}{2 c_W \Lambda}
\alpha_3^{NC} \; \bar{t} \gamma^5 t \; \partial^\mu Z_\mu
\nonumber \\
&+& \frac{g^2}{2 \sqrt{2} \Lambda c_W} \Biggl\{
\bar{t}~ \Bigl [\alpha_{1L}^{CC} (1-\gamma^5) + \alpha_{1R}^{CC}
(1+\gamma^5) \Bigr]b\; 
 W_\mu^+ Z^\mu 
\nonumber \\
 & + &\bar{b} \Bigl [\alpha_{1L}^{CC} (1+\gamma^5) +
 \alpha_{1R}^{CC}(1-\gamma^5)\Bigr]t \; 
 W^-_\mu Z^\mu \Biggr\} 
\label{lags} \\
 & + & i \frac{g}{2 \sqrt{2} \Lambda } \Biggl\{
 \bar{t}~ \Bigl [\alpha_{2L}^{CC} (1-\gamma^5) + \alpha_{2R}^{CC}
(1+\gamma^5) \Bigr]b \;
\left(\partial^\mu W_\mu^+ + i e A^\mu W_\mu^+\right) 
\nonumber \\ 
& - & \bar{b}~ \Bigl [\alpha_{2L}^{CC} (1+\gamma^5) + \alpha_{2R}^{CC}
(1-\gamma^5) \Bigr]t \;
\left(\partial^\mu W_\mu^- - i e A^\mu W_\mu^-\right)\Biggr\}\; . 
\nonumber
\end{eqnarray}
The Lagrangian containing vectorial current is given by,
\begin{eqnarray}
{\cal L}_V & = & i ~\frac{ g}{ 2 c_W}~ 
\gamma^{NC} ~\bar{t}  ~(\tilde D_\mu t) ~ Z^\mu 
 - i ~\frac{ g}{ 2 c_W}~ 
\gamma^{NC} \overline {(\tilde D_\mu t)} 
 ~t ~Z^\mu
\nonumber \\
& + &i ~\frac{ g}{ 2 \sqrt{2}}~ 
\bar{t} \left [ \gamma_L^{CC} (1-\gamma^5) + \gamma_R^{CC} 
(1+\gamma^5)\right]~
(\tilde D_\mu b) ~ W^{+\mu} 
\label{lagv} \\
& - & i ~\frac{ g}{ 4 c_W}~ 
\overline{(\tilde D_\mu b)} 
\left [ \gamma_L^{CC} (1+\gamma^5) + \gamma_R^{CC} 
(1-\gamma^5)\right]~t ~
 W^{-\mu} \; . \nonumber 
\end{eqnarray}
Finally, the piece involving a tensorial structure is,
\begin{eqnarray}
{\cal L}_T & = & \frac{1}{4 \Lambda}
\Biggl[ \; \bar{t} \; \sigma^{\mu\nu} \; t \; \Bigl 
( \beta_1^{NC} e F_{\mu\nu} + \beta_2^{NC} \frac{g}{c_W} Z_{\mu\nu}
+4 i g^2  \beta_3^{NC} W_\mu^+ W_\nu^{-} \Bigr) \Biggr] 
\nonumber \\
& + & \frac{g}{2 \sqrt{2} \Lambda} 
\Biggl\{
\bar{t} \;\sigma^{\mu\nu}\;\Bigl
[\beta_{1L}^{CC}(1-\gamma^5)+\beta_{1R}^{CC}
(1+\gamma^5) \Bigr] b \;
\Bigl[W^+_{\mu\nu}+i e\left(A_\mu W_\nu^+-A_\nu W_\mu^+\right) \Bigr]
\nonumber \\
& + &
\bar{b} \;\sigma^{\mu\nu} \;\Bigl
[\beta_{1L}^{CC}(1+\gamma^5)+\beta_{1R}^{CC}
(1-\gamma^5) \Bigr] t \;
\Bigl[W^-_{\mu\nu}-i e\left(A_\mu W^-_\nu-A_\nu W^-_\mu\right) \Bigr]
\label{lagt} \\
& + &i \frac{g}{c_W}~
\bar{t} \;\sigma^{\mu\nu} \;\Bigl
[\beta_{2L}^{CC}(1-\gamma^5)+\beta_{2R}^{CC}
(1+\gamma^5) \Bigr]b\;
\left(Z_\mu W_\nu^+-Z_\nu W_\mu^+\right)
\nonumber \\
& - &i\frac{g}{c_W}~
\bar{b}\; \sigma^{\mu\nu} \;\Bigl
[\beta_{2L}^{CC}(1+\gamma^5)+\beta_{2R}^{CC}
(1-\gamma^5) \Bigr] t  \;
\left(Z_\mu W^-_\nu-Z_\nu W^-_\mu\right)\; 
 \Biggr\} \; . \nonumber
\end{eqnarray}
The couplings constants $\alpha$'s, $\beta$'s and $\gamma$'s are
linear combinations of the $a$'s, $b$'s and $c$'s in Eqs.\
(\ref{nc}) to (\ref{cc}). In writing the interactions
(\ref{lags}) and (\ref{lagt}), the coupling constants were
defined in such a way that we have a factor $g/(2c_W)$ per $Z$
boson, $g/\sqrt{2}$ per $W^\pm$, and a factor $e$ per photon.
Similar interactions were obtained in Ref.\ \cite{yuan2} and for
a linearly realized symmetry group, in Ref.\ \cite{gouna}.

As an example of the above anomalous couplings, we show their
couplings for the SM with a heavy Higgs boson integrated out. In
this case, we can perform the matching between the full theory
and the effective Lagrangian \cite{dittmaier}. For instance, if
we concentrate on the non-decoupling effects, the leading
contributions come at one-loop  order \cite{foot20}. Setting
$m_b=0$ and keeping only the leading terms of the order $m_t
\log(M_H^2)$, we find that only the first two effective operators
of Eq.\ (\ref{lags}) are generated with coefficients,
\[
\alpha_1^{NC}=\alpha_2^{NC}=
\frac{g^2 m_t \Lambda}{16\pi^2 M_W^2}
\, \log\frac{M_H^2}{m_t^2} \; .
\]

\section{Results for the $b \to s \gamma$ and $b \to s \ell^+ \ell^-$
transitions}

For the $b\to s\gamma$ and $b\to s\ell^+\ell^-$  transitions it is 
useful to cast the contributions of the dimension-four 
and dimension-five anomalous couplings 
as shifts in the matching conditions at $M_W$ for the 
Wilson coefficient functions in the weak effective 
hamiltonian  
\begin{equation}
H_{\rm eff.}=-\frac{4G_F}{\sqrt{2}}V_{tb}^*V_{ts}\sum_{i=1}^{10} 
C_i(\mu)O_i(\mu),
\label{heff}
\end{equation}
with the operator basis defined in Ref.~\cite{heff}.
Of interest in our analysis are the electromagnetic penguin operator
\begin{equation}
{\cal O}_7 = \frac{e}{16\pi^2}m_b\; 
(\bar{s}_L\sigma_{\mu\nu}b_R)\;F^{\mu\nu}~,
\label{o7}
\end{equation}
and the four-fermion operators corresponding to the vector and
axial-vector couplings to leptons, 
\begin{equation}
{\cal O}_9 = \frac{e^2}{16\pi^2}\;(\bar{s}_L\gamma_\mu b_L) 
(\bar{\ell}\gamma^\mu\ell) \; ,
\label{o9}
\end{equation}
and
\begin{equation}
{\cal O}_{10} = \frac{e^2}{16\pi^2}\;(\bar{s}_L\gamma_\mu b_L)
(\bar{\ell}\gamma^\mu\gamma_5\ell)~.
\label{o10}
\end{equation}
The operators above are already present in the SM. In addition, 
the dimension-five anomalous 
couplings generate the operators 
\begin{eqnarray}
{\cal O}_{11} &=& \frac{e^2}{16\pi^2} \frac{m_b}{M_Z^2} \; 
[\bar{s}_L\sigma_{\mu\nu} (i Q^\nu)  b_R]\;
(\bar{\ell}\gamma^\mu\gamma_5\ell)~,
\label{o11}\\
{\cal O}_{12} &=& \frac{e^2}{16\pi^2} \frac{m_b}{M_Z^2} \; 
[\bar{s}_L\sigma_{\mu\nu} (i Q^\nu) b_R]\;(\bar{\ell}\gamma^\mu \ell)~.
\label{o12}
\end{eqnarray}
However, these new operators will not lead to important effects
as will see below, due to the fact that they are further
suppressed  by the weak scale. 

The anomalous couplings of Eq.~(\ref{lfer}), (\ref{lags}),
(\ref{lagv}) and  (\ref{lagt}) will induce shifts in the Wilson
coefficient functions $C_i(\mu)$  at the matching scale, which we
take to be $\mu=M_W$.  We make use of the next-to-leading order
calculation of the Wilson coefficients as described in
Ref.~\cite{nlo}. 

\subsection{Effects of the Dimension--four operators}

The dimension-four operators defined in Eq.~(\ref{lfer}) induce new
contributions  to the $b\to s\gamma$ 
and $b\to s Z$ loops 
as well as the box diagram. They appear in the effective Hamiltonian
formulation as shifts of the Wilson coefficients $C_7(M_W)$,
$C_9(M_W)$ and $C_{10}(M_W)$. The contributions 
from the $b\to s\gamma$ loops to $C_7(M_W)$ are:
\begin{eqnarray}
\delta C_7&=& -\frac{m_t}{m_b} C_R \left\{
\frac{1}{12 (x - 1)^2}  \; (5 x^2 - 31 x + 20) 
\right.\nonumber \\
&& \left. +\frac{1}{2 (x - 1)^3}  \; x (3 x - 2) \log\left(x\right) 
\right\}\label{C74} \\
&& + C_L \left\{\frac{1}{4 (x - 1)^4}  \; x^2 (3 x - 2) 
\log\left(x\right)\right. \nonumber \\
&& \left.-\frac{1}{24 (x - 1)^3} \; x (8 x^2 + 5 x - 7) 
\right\}\; , \nonumber 
\end{eqnarray}
where we have defined the dimensionless quantity $x =
m_t^2/M_W^2$.  We should notice that the above result is finite,
{\it i.e.\/} independent of $\Lambda$, and agrees with the
previous result in the literature \cite{fuji}. On the other hand,
the result for all other operators is not finite and in this case
we have kept only the leading non--analytic, {\it i.e.\/}
logarithmic, dependence on  the new physics scale $\Lambda$. In
this way, for $C_9(M_W)$ we have,
\begin{equation}
\delta C_9^\gamma = - \frac{1}{12}   C_L \; (3x - 16) 
\log\left(\frac{\Lambda^2}{M_W^2}\right)\; .
\label{C9g4}
\end{equation}

The corrections arising from $b\to s Z$ loops to $C_9(M_W)$ and 
$C_{10}(M_W)$ are:
\begin{equation}
\delta C_{10}^Z = \frac{-1}{1 - 4 s_W^2} \delta C_{9}^Z =  
\frac{1}{16 s_W^2 } (4 N_L^t - N_R^t +
C_L) \; x \; \log\left(\frac{\Lambda^2}{M_W^2}\right)  \; ,
\label{C9Z4} 
\end{equation}
while box loops contributions can be written as:
\begin{equation}
\delta C_9^{\mbox{box}}= -\delta C_{10}^{\mbox{box}}= 
\frac{1}{16 s_W^2}   C_L \; 
(x - 16)  \log\left(\frac{\Lambda^2}{M_W^2}\right)\; .
\label{C9b4} 
\end{equation}

The measured $b\to s\gamma$ branching ratio imposes a stringent
bound on $C_R$ as its contribution to  (\ref{C74}) is enhanced by
the factor $m_t/m_b$. This has been discussed in the
literature~\cite{fuji}, where the obtained bounds on $C_R$  :
$-0.05<C_R<0.01$.  However, in the spirit of naturalness in a
strongly  coupled theory it is hard to justify such small values
for this coefficient unless there is a symmetry protecting this
term. In this case, chiral symmetry is violated by $C_R$, which
is then forced to be very small. In order to see  this, we notice
that $C_R$ would contribute to the renormalization of the
$b$-quark line with a term which does not vanish in the $m_b\to
0$ limit
\begin{equation}
\Sigma(m_b)=\frac{g^2}{32 \pi^2} \, C_R \, m_t\, (x-4)\,
\log\left(\frac{\Lambda^2}{M_W^2}\right)\;.
\end{equation}
Thus we are inclined to redefine this coefficient by
defining $\hat C_R$ as  
\begin{equation}
C_R = \frac{m_b}{\sqrt{2}v}\hat C_R , 
\label{crst}
\end{equation}
where $v=246$~GeV. With this redefinition, the contributions of
$\hat C_R$  to the $b$-quark mass vanish in the chiral limit. The
rescaled bounds on  $\hat C_R$ are now ${\cal O}(1)$, thus allowing
for more natural values of  this coefficient. 

\begin{figure} 
\centerline{
\epsfig{file=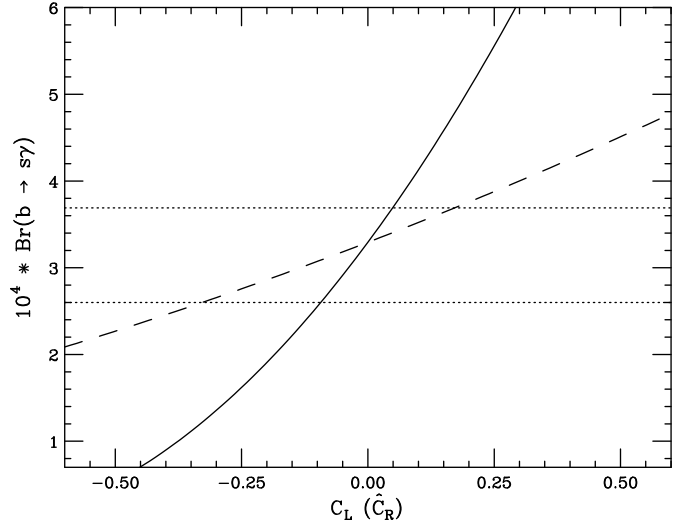,height=3.5in,angle=90}
}
\caption{The $b\to s\gamma$ branching ratio vs. $\hat C_R$ (solid)
and $C_L$ (dashed).  
The horizontal band corresponds to the $1\sigma$ interval from 
the latest CLEO result~\protect\cite{cleo}. }
\label{bsgd4}
\end{figure}

In Fig.~\ref{bsgd4}
we plot the $b\to s\gamma$ branching  fraction as a function of
$\hat C_R$. We also include the effect of $C_L$, which  is now
comparable for similar values of the coefficients. 
The horizontal band corresponds to the latest CLEO
result~\cite{cleo} $Br(b\to s\gamma)=(3.15\pm 0.35\pm 0.32\pm
0.26)\times 10^{-4}$, where we take a $1\sigma$ interval after
adding the statistical, systematic and  model-dependence errors
in quadrature.

On the other hand, the effect in $b\to s\ell^+\ell^-$ is
dominated by the coefficients $C_L$, $N^t_L$ and $N^t_R$ 
in Eqs. (\ref{C9g4}), (\ref{C9Z4}), and (\ref{C9b4}).  
In principle, these coefficients are constrained  by electroweak
precision measurements, most notably $\epsilon_1=\Delta\rho=\alpha T$ and
$R_b$ \cite{yuan}: 
\begin{eqnarray}
\epsilon_1&=&\frac{G_F}{2 \sqrt{2} \pi^2}3 m_t^2\left(-N^t_L+N^t_R+C_L\right)
\;\log\left(\frac{\Lambda^2}{M_W^2}\right) \nonumber\\
\epsilon_b&=&\frac{G_F}{2 \sqrt{2} \pi^2}3 m_t^2
\left(-\frac{1}{4} N^t_R + N^t_L \right)\;
\log\left(\frac{\Lambda^2}{M_W^2}\right)\; .
\end{eqnarray}
In general, the bounds
obtained on a particular coupling from electroweak observables
strongly depend on assumptions about the other
couplings. For instace, enforcing custodial isospin symmetry in order to avoid
the strong constraints from $T$ will imply that $N^t_L=C_L$ and $N^t_R=0$. 
On the other hand if $C_L=0$, then the combination
$(N_L^t-N_R^t)$ is strongly constrained since it breaks custodial isospin
symmetry and contributes to $T$. Impossing $C_L = N_L^t$, then  
$ N_R^t < 0.02$~\cite{peccei,bdjt} since it is the only linear
source of isospin breaking.

We study here three cases in which the stringent constraints from 
electroweak observables can be evaded. 

\begin{figure} 
\centerline{
\epsfig{file=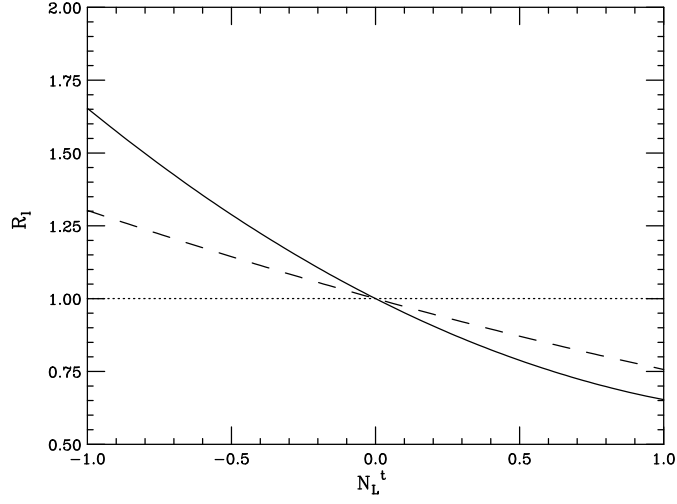,height=3.5in,angle=90}
}
\caption{The $b\to s\ell^+\ell^-$ branching ratio vs. $C_L=N_L^t$
(solid) and  $N_L^t=N_R^t$ (dashed). 
}
\label{fig2}
\end{figure}

{\em i) $C_L\approx N_L^t$}. In this case the contributions of
$C_L$ and $N_L^t$ to  the $T$ parameter cancel, leaving $N_R^t$
as the only seriously constrained quantity. However, $R_b$ still
gives the bound $-0.03<N_L^t<0.15$ (for $\Lambda=1$ TeV).   
In Fig.~\ref{fig2}  we plot the
$b\to s\ell^+\ell^-$ branching ratio, normalized to  the SM
expectation, as a function of $C_L=N_L^t$ (solid line).  From
this plot it can be seen that, when  incorporating the $R_b$
constraint, the effect in $b\to s\ell^+\ell^-$ is bound to be
smaller than roughly a $10\%$ deviation. 

{\em ii) $N_L^t\approx N_R^t$}. In this scenario the measurement
of the $T$ parameter  greatly constrains $C_L$, which prompts us
to take this coefficient as equal to zero  in this portion of the
analysis. The dashed line in Fig.~\ref{fig2} corresponds to the
effect of $N_L^t=N_R^t$ in the dilepton branching fraction. As we
can see, the effect in this decay is rather small. 

{\em iii)} Finally and for completeness, we consider the case
$N_L^t\approx N_R^t/4$.  With this approximate relation these
two coefficients cancel in $R_b$ leaving no  sizeable effect
from the dimension-four lagrangian (\ref{lfer}) in this quantity.
However, this relation leads to a potentially large contribution
to the $T$ parameter proportional to $(C_L+3N_R^t/4)$. When this
bound is incorporated, the effect in  $b\to s\ell^+\ell^-$
branching ratio is constrained to be below $15\%$.

In sum, we have seen that the leading effects of the
dimension--four operators  in rare $B$ decays 
are given by 
$\hat C_R$ in $b\to s\gamma$, and the  effects in $b\to
s\ell^+\ell^-$ due to $C_L$, $N_L^t$ and $N_R^t$  are 
below $15\%$ deviations once the constraints from
electroweak  precision measurements are considered. This
distinction comes from the fact that $Z$-pole quantities  are not
significantly sensitive to $\hat C_R$. 
The
effects of $\hat C_R$ in $b\to s\ell^+\ell^-$ can be significant, but
$b\to s\gamma$  is considerably more sensitive to this
parameter.

\subsection{Effects of the Dimension--five Operators}

Although in principle dimension-five operators are considered
sub-leading  with respect to the operators in Eq.~(\ref{lfer})
due to the additional suppression by the high energy scale
$\Lambda$, they can still induce large deviations in  both
electroweak observables and FCNC processes. In Ref.~\cite{ewd5}
bounds on the  coefficients of dimension-five operators were
derived from data at the Z--pole. Here we consider the effect of
these operators in $b\to s\gamma$ and $b\to s\ell^+\ell^-$.
They  induce new contributions  to the $b\to s\gamma$  and $b\to
s Z$ loops  as well as the box diagram. They appear in the
effective Hamiltonian formulation as shifts of the Wilson
coefficients $C_7(M_W)$, $C_9(M_W)$ and $C_{10}(M_W)$,
$C_{11}(M_W)$, and $C_{12}(M_W)$.

The contribution from the $b\to s\gamma$ loops  to these coefficients 
are:
\begin{equation}
\delta C_7 = - \frac{1}{12}  
\frac{M_W^2}{m_b \Lambda} [4 \alpha_{2R}^{CC} x - 4 \beta_{1R}^{CC}
(3 x - 7)  + \gamma_{R}^{CC} (x + 2) ]
\log\left(\frac{\Lambda^2}{M_W^2}\right) \; ,
\label{C75}
\end{equation}
and 
\begin{equation}
\delta C_{12}^\gamma= \frac{1}{24} \frac{M_Z^2}{m_b\Lambda} \Big\{
\Big[ 3 ( 2 \alpha_{2R}^{CC}  - 
\gamma_{R}^{CC}) x  + 4 (3 \alpha_{2R}^{CC} 
- 6 \beta_{1R}^{CC} + \gamma_{R}^{CC}) \Big]  \Big\}
\log\left(\frac{\Lambda^2}{M_W^2}\right) \; . 
\label{c12g}
\end{equation}

The corrections arising from $b\to s Z$ loops to the different
coefficients are:
\begin{eqnarray}
\delta C_{10}^Z &=& \frac{-1}{1 - 4 s_W^2} \delta C_{9}^Z =  
- \frac{1}{96 s_W^2} \frac{m_t}{\Lambda}
\nonumber \\ 
&\times& \Big[ 3 (2\alpha_{1L}^{CC} + 12  \alpha_{2L}^{CC} + 
12 \beta_{2L}^{CC} - \gamma_{L}^{CC} + 8 \gamma^{NC}) x
- 18 c_W^2 (\alpha_{2L}^{CC}  + 2 \beta_{1L}^{CC}) x \nonumber \\
& +&
2  s_W^2 (\gamma_{L}^{CC} - 12 \alpha_{2L}^{CC} ) x
- 9(2\alpha_{1L}^{CC} - 4 \beta_{1L}^{CC} + 12
\beta_{2L}^{CC} - \gamma_{L}^{CC} - 4 \gamma^{NC}) \\ 
&-& 18 c_W^2 (\alpha_{2L}^{CC}  - 2 \beta_{1L}^{CC}) - 
6 s_W^2 (\gamma_{L}^{CC} + 12 \beta_{1L}^{CC}) 
\Big]
\log\left(\frac{\Lambda^2}{M_W^2}\right) \; , \nonumber
\end{eqnarray}
and 
\begin{eqnarray}
\delta C_{11}^Z &=& \frac{-1}{1 - 4 s_W^2} \delta C_{12}^Z =  
- \frac{1}{48 s_W^2}  \frac{M_Z^2}{m_b\Lambda}
\nonumber \\
&\times& \Bigl[ 
3 (\alpha_{1R}^{CC}  + \alpha_{2R}^{CC} - 2 \beta_{2R}^{CC}) x 
+ 3 c_W^2 (\alpha_{2R}^{CC} - 6 \beta_{1R}^{CC} + \gamma_{R}^{CC}) x  
\label{c11z} \\
 &-& 2 s_W^2 (4 \alpha_{2R}^{CC} - \gamma_{R}^{CC}) x 
+ 3 (2 \alpha_{1R}^{CC} + 8 \beta_{1R}^{CC} - 4 \beta_{2R}^{CC}
- \gamma_{R}^{CC})
\nonumber \\
&+& 6 c_W^2 (\alpha_{2R}^{CC} + 6 \beta_{1R}^{CC} + \gamma_{R}^{CC})
- 4 s_W^2 (8  \beta_{1R}^{CC} - \gamma_{R}^{CC})
\Bigr] \nonumber
\log\left(\frac{\Lambda^2}{M_W^2}\right)\; .  \nonumber
\end{eqnarray}
The box loops contributions can be written as:
\begin{equation}
\delta C_9^{\mbox{box}}= -\delta C_{10}^{\mbox{box}}= 
- \frac{1}{96 s_W^2} 
\frac{m_t}{\Lambda} 
\Bigl[ 4 \alpha_{2L}^{CC} (2 x - 5) - 36 \beta_{1L}^{CC}
- \gamma_{L}^{CC} (x - 1) \Bigr]
\log\left(\frac{\Lambda^2}{M_W^2}\right)\; ,
\end{equation}
and 
\begin{equation}
\delta C_{12}^{\mbox{box}}= -\delta C_{11}^{\mbox{box}}= 
-\frac{1}{16 s_W^2} 
\frac{M_Z^2}{m_b\Lambda}
\Bigl[ 2 \alpha_{2R}^{CC} x  - 4 \beta_{1R}^{CC} 
- \gamma_{R}^{CC} (x - 2) \Bigr]
\log\left(\frac{\Lambda^2}{M_W^2}\right)  \; .
\label{c11box}
\end{equation}

\begin{figure} 
\centerline{
\epsfig{file=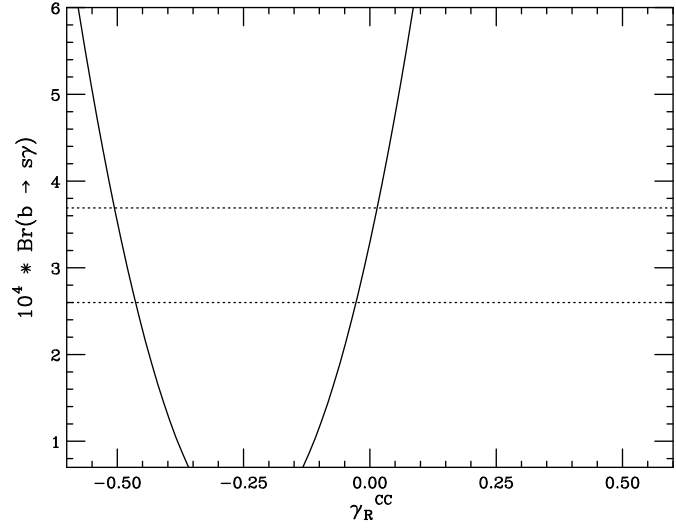,height=3.5in,angle=90}
}
\caption{The $b\to s\gamma$ branching ratio vs. $\gamma_R^{CC}$. 
}
\label{grcc}
\end{figure}

\begin{figure} 
\centerline{
\epsfig{file=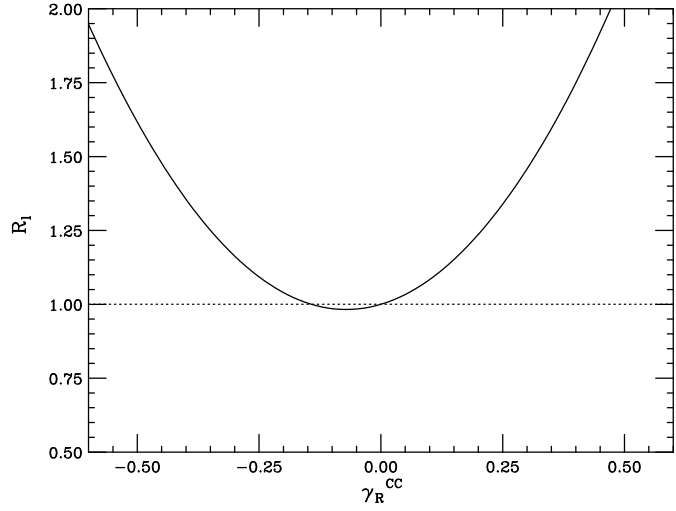,height=3.5in,angle=90}
}
\caption{The $b\to s\ell^+\ell^-$ branching ratio (normalized to 
the SM prediction) vs. $\gamma_R^{CC}$. 
}
\label{bslgrcc}
\end{figure}

In order to quantify 
the effect of the operators of Eqs.~(\ref{lags}),~(\ref{lagv})
and (\ref{lagt}) we classify them into two groups: the
right--handed and left--handed couplings.  The RH couplings that
are of interest due to their potential effects are
$\alpha_{2R}^{CC}$, $\beta_{1R}^{CC}$ and $\gamma_R^{CC}$. Just
as for the dimension-four coefficient $C_R$, the effects of these
coefficients can be very large in operators such as ${\cal O}_7$
generating important deviations in $b\to s\gamma$, as seen in
Eq.~(\ref{C75}). This is particularly so since their contributions
are unsuppressed by $m_b$. However, and as we noted for $C_R$,
the coefficients $\alpha_{2R}^{CC}$ and $\beta_{1R}^{CC}$ are
chirally unsuppressed in Eq.~(\ref{lags}), which results in an
unnatural renormalization of the $b$-quark mass. Thus again we
would argue that  these coefficients should be rescaled by the
factor $m_b/v$, which makes  their effect on the operator $O_7$
negligible. On the other hand, this is not the case with the
coefficient $\gamma_R^{CC}$ since the chiral suppression  is
already present in the accompanying  operator in 
Eq.~(\ref{lagv}). Because of this its contribution to the 
to the renormalization of the $b$-quark line vanishes in 
the $m_b\to 0$ limit
\begin{equation}
\Sigma(m_b)=
\frac{-3 g^2}{128 \pi^2 \Lambda} \, \gamma_R^{CC} \, m_b^2\, (x+1)\,
\log\left(\frac{\Lambda^2}{M_W^2}\right)\;.
\end{equation}
We will then concentrate on the effects of $\gamma_R^{CC}$  among
the RH couplings. The contribution of $\gamma_R^{CC}$ to the
penguin operator gives rise to  a deviation of the $b\to s\gamma
$ branching ratio from its SM expectation. In Fig.~\ref{grcc} we
plot this branching fraction as a function of  this coefficient.
This  measurement is the most constraining bound on these type of
operators.  It can be seen that even for rather small values of
$\gamma_R^{CC}$ there could be considerable deviations from the
SM expectations.  On the other hand, the effect is less dramatic
in $b\to s\ell^+\ell^-$,  as shown in Fig.~\ref{bslgrcc}, where
an observable deviation from the  SM will result only if
$\gamma_R^{CC}$ is large enough to dominate  the $b\to s\gamma$
branching ratio.

The effects of the new operators ${\cal O}_{11}$ and ${\cal
O}_{12}$  are negligible. Although the presence of $m_b$ in the
denominators in  Eq.~(\ref{c12g}), (\ref{c11z}) and
(\ref{c11box}) suggests the possibility of an enhancement, this
is not enough. This is obviously true for the  coefficients
$\alpha_{2R}^{CC}$ and $\beta_{1R}^{CC}$, which as we argue
above should be proportional to $m_b/v$. But even when
considering  $\gamma_{R}^{CC}$, the effect is suppressed by an
effective scale  given by $m_b\Lambda/M_Z\simeq 55~$GeV, which
should be compared with the  typical momentum transfers in $B$
decays.  

\begin{figure} 
\centerline{
\epsfig{file=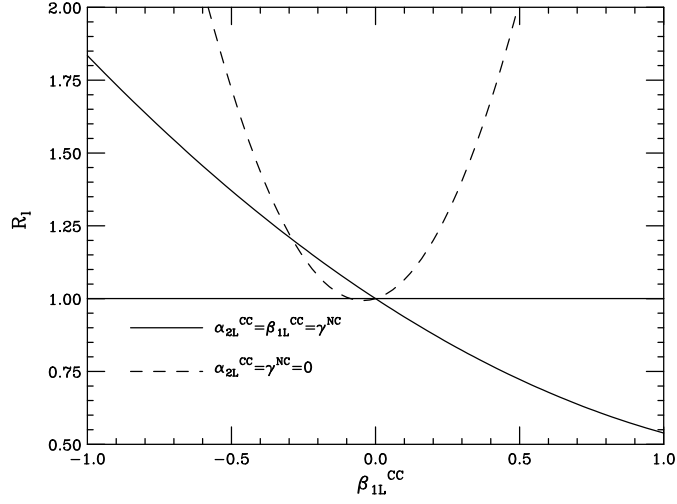,height=3.5in,angle=90}
}
\caption{The $b\to s\ell^+\ell^-$ branching ratio (normalized to 
the SM prediction) vs. $\beta_{1L}^{CC}$, for   
$\alpha_{2L}^{CC}=\beta_{1L}^{CC}=\gamma^{NC}$ (solid line) 
and $\alpha_{2L}^{CC}=\gamma^{NC}=0$ (dashed line). 
}
\label{lh}
\end{figure}

The remaining group of coefficients we dubbed left-handed
includes $\alpha_{1L}^{CC}$, $\alpha_{2L}^{CC}$,
$\beta_{1L}^{CC}$,  $\beta_{2L}^{CC}$, $\gamma_L^{CC}$ plus
$\gamma^{NC}$ which actually is the coefficient of a vector
operator, but since is not chirally suppressed is included with
the LH in this part of the analysis. These operators affect
mainly the $b\to s\ell^+\ell^-$ rates.  Thus it is possible to
imagine that the underlying new physics  preserves chiral
symmetry at the same time that does not generate a large value of
$\gamma_R^{CC}$, resulting in no deviations in $b\to s\gamma$;
but that  the effects of the new interactions  give rise to large
effects in the dilepton modes. Although  these have not been
observed yet the experimental sensitivity is very  close to the
SM predictions and it will reach them in the near future.
The leading effects in $b\to s\ell^+\ell^-$  come from the
coefficients $\alpha_{2L}^{CC}$, $\beta_{1L}^{CC}$ and
$\gamma^{NC}$. For simplicity we only consider these and plot in
Fig.~\ref{lh} the branching ratio, normalized to the SM one, for
two  cases: $\alpha_{2L}^{CC}=\beta_{1L}^{CC}=\gamma^{NC}$ (solid
line)  and $\alpha_{2L}^{CC}=\gamma^{NC}=0$ (dashed line). 
>From Fig.~\ref{lh} it is apparent that cancellations occur when
the three  coefficients are similar. The effect of considering only
$\beta_{1L}^{CC}$ shows than even larger effects are possible. In
any case, sizeable deviations in $b\to s\ell^+\ell^-$ are possible
even in the absence of effects in $b\to s\gamma$.

\section{Discussion}

Processes involving FCNC transitions in $B$ and $K$ decays are a
crucial complement to precision electroweak observables, when
constraining the physics of the EWSB sector.   In this paper, we
have considered the effects of anomalous couplings of third
generation quarks to the $W$ and $Z$ gauge bosons. We computed
the effects of   all possible dimension-five operators in $B$
FCNC transitions such as $b\to s\gamma$ and $b\to s\ell^+\ell^-$.
For completeness, we have also also presented the analysis of the
dimension-four operators.  

We have shown that with the natural assumption of chiral
symmetry,  in fact enforcing vanishing fermion mass
renormalization in the chiral limit, the effects of the
dimension-four operators with coefficient  $C_R$  for $b\to
s\gamma$, are not as dramatic as found Ref.~\cite{fuji},   and
somehow smaller than those of $C_L$, which can produce important
deviations in the branching ratio that could be resolved  in the
next round of experiments at $B$ factories.

The  effects in $b\to s\ell^+\ell^+$ due to  $C_L$, $N_L^t$ and
$N_R^t$  are below $15\%$ deviations once the constraints from
electroweak  precision measurements are taken into account. 

On the other hand, we have found that the dimension-five operator
with coefficient $\gamma^{CC}_R$, for which no additional chiral
suppression is expected,  can  give rise to an observable
deviation of the $b\to s\gamma$  branching ratio from its SM
expectation even for rather small values of $\gamma_R^{CC}$. Left
handed operators, on the other hand, affect mainly the $b\to
s\ell^+\ell^-$ rates and we have  illustrated that in several
scenarios   sizeable deviations in $b\to s\ell^+\ell^-$ are
possible even in the absence of effects in $b\to s\gamma$. 

The dimension-five operators, just as in the case of  the more
studied dimension-four operators, can be generated at  high
energies scales by the presence of new particles and/or
interactions. For instance, as a simple example, a heavy scalar
sector with both charged and  neutral states, would give
contributions to many of the coefficients of the Lagrangian in
Eq.~(\ref{lags}). Richer dynamics at the TeV scale might
generate also some of the  vector and/or tensor couplings of
Eq.~(\ref{lagv}) and~(\ref{lagt}). 

The $e^+e^-$ $B$ factories at Cornell, KEK and SLAC are expected
to reach better measurements of the $b\to s\gamma$ branching
ratio,  which will largely constrain the dimension-five
coefficient  $\gamma_R^{CC}$ and to a lesser extent the
dimension-four coefficients $C_L$ and $\hat C_R$.  Furthermore,
these experiments as well as those at the Fermilab  Tevatron,
will reach the SM sensitivity for the $b\to s\ell^+\ell^-$
branching fraction. The present analysis, together with previous
ones  addressing the effects of other anomalous higher
dimensional operators in these decay modes, will enable us to
interpret a possible pattern of deviations from the SM and
perhaps point to its dynamical origin.

\vskip 1.0cm
\noindent
{\bf Acknowledgments}

\noindent
We thank O.\ J.\ P.\ \'Eboli for a critical reading of the
manuscript. M.\ C.\ Gonzalez--Garcia is grateful to the Instituto
de F\'{\i}sica Te\'orica from Universidade Estadual Paulista for
its kind hospitality. G.\ Burdman acknowledges the hospitality of
the Enrico Fermi Institute at the University of Chicago, where
part of this work was completed. This work was supported by the U.S.\
Department of Energy under Grant No.~DE-FG02-95ER40896 and the
University of Wisconsin Research Committee with funds granted by
the Wisconsin Alumni Research Foundation, by Conselho Nacional de
Desenvolvimento Cient\'{\i}fico e Tecnol\'ogico (CNPq), by
Funda\c{c}\~ao de Amparo \`a Pesquisa do Estado de S\~ao Paulo
(FAPESP), by Programa de Apoio a N\'ucleos de Excel\^encia
(PRONEX), by CICYT (Spain) under grant AEN96-1718, by DGICYT
(Spain) under grants PB95-1077 and PB97-1261, and by the EEC
under the TMR contract ERBFMRX-CT96-0090.



\begin{thebibliography}{99}
\bibitem{foot20} The tree-level Higgs exchange goes like
$g^2m_tM_W/vM_H^2$ and although it could be numerically significant, 
formally decouples in the large $M_H$ limit.
\end{thebibliography}
\end{document}